\documentclass{aastex}
\usepackage{spr-astr-addons}
\usepackage{url}
\usepackage[dvips]{color}
\usepackage{epsfig}
\usepackage{amsmath}
\usepackage{graphicx}
\RequirePackage{color}

\begin{document}
\title{Effect of magnetic field on the rotating filamentary molecular clouds}
\author{P. Aghili}  \author{K. Kokabi}
\affil{School of Physics, Damghan University, Damghan, Iran.\\
kokabi@du.ac.ir}

\begin{abstract}
Purpose of this work is study evolution of magnetized rotating filamentary molecular cloud's. We will consider cylindrical symmetric filamentary molecular clouds at the early stage of evolution. For the first time we consider rotation of filamentary molecular in presence of an axial and azimuthal magnetic field without any assumption of density and magnetic functions. We show that in addition to decreasing of radial collapse velocity, the rotational velocity also affected by magnetic field. Existence of rotation yields to fragment of filament. Moreover, we show that the magnetic field have significant effect on the fragmentation of filamentary molecular clouds.
\end{abstract}
\section{Introduction}
Giant molecular clouds are one of the most important star-formation regions. Filamentary structures are present in many molecular clouds. They may be birthplace for stars or even many body systems. So, analyzing their evolution and dynamics help us to better understanding of star formation process. Physical processes like self-gravitation process, thermal process and magnetic field play the main roll in the star-formation process. (Larson 1985, Nakamura1995, Nakajima 1996, Shadmehri 2005, Van Loo et al. 2014). \\
Various magnetic field observed in the filamentary molecular cloud (Heiles 2000; Goldsmith et al. 2008). There are various configuration of magnetic field with respect to the filament such as parallel or orthogonal (Poidevin et al 2011; Vrba 1976). Value of magnetic field is also different at different place of order micro G (Opher et al. 2009; Ferriere 2009). Already, effect of magnetic field on the filamentary molecular cloud studied, some of them used a function for magnetic field, and some of them assumed special magnetic field axis (Nagasava 1987). Now our main goal is to study effect of magnetic field on the rotational filamentary molecular cloud without any assumption about function of magnetic field. Already (Khesali et al. (2014)) we studied effect of magnetic field on the filamentary molecular cloud without rotation and found no fragment. Now, we will show that effect of rotational motion is fragment of filament which also affected by magnetic field.\\
Considering various magnetic fields that observed in molecular clouds, result gained  for spherical symmetric collapse in the rotation by (Ulrich 1976). With generalizing those results to disk structure by (Cassen and Moosman 1981) and to isothermal crash (collapse-Decay) by (Terbey, Shu and Cassen 1984) the results were analyzing. (Mendoza, Tejeda and Nagel 2009 here after MTN). Generalized this idea with creation of a fluid continuous growing flow model for a definite gas cloud by considering outer region observations of filamentary cloud. We will study the effect of magnetic field on the other filament parameters like density, rotational velocity and etc. We will study cooling function changes and rotation effect on the molecular clouds.
\section{Main equations}
In this section, as regards that we have symmetry in our problem, we will consider the cloud as a long cylinder with symmetrical z axis, and considering magnetic field as  $B=B_{\varphi} \hat{\varphi}+B_{z}\hat{k}$ ( $B$ is function of radial distance and time, i.e. $B(r, t)$ ) considering Ideal gas and energy equations, we have main equations as:
\begin{equation}\label{1}
\frac{\partial\rho}{\partial t}+\nabla\cdot(\rho u)=0,
\end{equation}

\begin{eqnarray}\label{2}
\frac{\partial v_{r}}{\partial t}+v_{r}\frac{\partial v_{r}}{\partial r}+\frac{1}{\rho}\frac{\partial P}{\partial r}+\frac{\partial \psi}{\partial r}=\frac{v_{\varphi}^{2}}{r}\nonumber\\
-\frac{1}{4\pi \rho}\left[\frac{B_{\varphi}}{r}\frac{\partial}{\partial r}(rB_{\varphi})+B_{z}\frac{\partial B_{z}}{\partial r}\right],
\end{eqnarray}

\begin{equation}\label{3}
\frac{1}{r}\frac{\partial}{\partial r}(r \frac{\partial \psi}{\partial r})=4\pi G \rho
\end{equation}

\begin{equation}\label{4}
\frac{\partial B}{\partial t}=\nabla\times(v\times B)
\end{equation}

\begin{eqnarray}\label{5}
\frac{1}{\gamma-1}\left[\frac{\partial p}{\partial t}+v_{r}\frac{\partial p}{\partial r}\right]&+&\frac{\gamma}{\gamma-1}\frac{p}{r}\frac{\partial}{\partial r}(r v_{r})\nonumber\\
&+&A_{\nu}\rho^{2}T^{\nu}=0,
\end{eqnarray}

\begin{equation}\label{6}
\frac{\partial}{\partial t}(r v_{\varphi})+v_{r}\frac{\partial}{\partial r}(r v_{\varphi})=0,
\end{equation}

where$\rho$, $V_{r}$, $V_{\varphi}$, $\psi$, $P$ and $B$ are: gas density, radial velocity, azimuthal velocity, gravitational potential, pressure and magnetic field. Also $\nu$ and $A_{\nu}$ are constants. These constants can be determined by Spritzer (1978). All changes are depend on $R$ (the distance from cylinder axis) and time. For simplification, we will introduce dimensionless variables and then solve the problem using the following translations

\begin{eqnarray}\label{7}
\rho\rightarrow\rho\hat{\rho}, \hspace{2mm}t\rightarrow t\hat{t}, \hspace{2mm} v\rightarrow v\hat{v}, \hspace{2mm}r\rightarrow r\hat{r},\nonumber\\
p\rightarrow p\hat{p}, \hspace{2mm}B\rightarrow B\hat{B}, \hspace{2mm}\psi\rightarrow\psi\hat{\psi},
\end{eqnarray}
where

\begin{eqnarray}\label{8}
\hat{\rho}&=&\rho_{0}, \hspace{2mm} \hat{P}=P_{0}, \hspace{2mm} \hat{\psi}=\frac{P_{0}}{\rho_{0}}, \hspace{2mm} \hat{r}=\sqrt{\frac{P_{0}}{4\pi G \rho_{0}}},\nonumber\\
\hat{t}&=&\frac{P_{0}}{A_{\nu}\rho_{0}^{2}T_{0}^{\nu}}, \hspace{2mm} \hat{v}=\frac{\hat{r}}{\hat{t}}, \hspace{2mm} \hat{B}=\sqrt{4\pi P_{0}}.
\end{eqnarray}

We choose the physical scale for length and time which are 1PC and $10^{6}$ yr, respectively. Thus, the gravitational constant is $G=4.48\times10^{-3}$. In this manner, the temperature and density units are 130 K and $6.67\times 10^{-23} \frac{gr}{Cm^{3}}$, respectively. And thee unit of magnetic field is chosen equal to 2.85 $\mu G$.

\section{Self-similar solution}
By considering the modified equations (7), we will replace dimensionless functions of $(r,t)$ with the function of $(\eta, t)$ for self-similar variable $\eta=\frac{r}{t^{n}}$. We use the following forms of the physical variables,

\begin{eqnarray}\label{9}
\rho(r, t)&=&R(\eta)t^{\varepsilon_{1}},\nonumber\\
v_{r}(r, t)&=&v_{r}(\eta)t^{\varepsilon_{2}},\nonumber\\
\psi(r, t)&=&\psi(\eta)t^{\varepsilon_{3}},\nonumber\\
B_{z}(r, t)&=&B_z(\eta)t^{\varepsilon_{4}},\nonumber\\
B_{\varphi}(r, t)&=&B_{\varphi}(\eta)t^{\varepsilon_{5}},\nonumber\\
P(r, t)&=&P(\eta)t^{\varepsilon_{6}},\nonumber\\
v_{\varphi}(r, t)&=&v_{\varphi}(\eta)t^{\varepsilon_{7}},
\end{eqnarray}
By equilibrating the time powers, local equations will be obtained as follows,
\begin{equation}\label{10}
\varepsilon_{1}R(\eta)-n\eta\frac{dR(\eta)}{d\eta}+\frac{1}{\eta}\frac{d}{d\eta}(\eta R(\eta)v_{r}(\eta))=0,
\end{equation}

\begin{eqnarray}\label{11}
\varepsilon_{2}v_{r}(\eta)&-&n\eta\frac{dv_{r}(\eta)}{d\eta}+v_{r}(\eta)\frac{dv_{r}(\eta)}{d\eta}+\frac{1}{R(\eta)}\frac{dP(\eta)}{d\eta}\nonumber\\
+\frac{d\psi(\eta)}{d\eta}&=&-\frac{B_{z}(\eta)}{R(\eta)}\frac{dB_{z}(\eta)}{d\eta}-\frac{B_{\varphi}(\eta)}{\eta R(\eta)}\frac{d(\eta B_{\varphi}(\eta))}{d\eta}
+\frac{v_{\varphi}^{2}}{\eta},\nonumber\\
\end{eqnarray}

\begin{equation}\label{12}
(v_{r}(\eta)+\varepsilon_{7}\eta)v_{\varphi}(\eta)+(v_{r}(\eta)-n\eta)\eta\frac{d v_{\varphi}(\eta)}{d\eta}=0,
\end{equation}

\begin{equation}\label{13}
\frac{1}{\eta}\frac{d}{d\eta}(\eta \frac{d \psi(\eta)}{d\eta})=R(\eta),
\end{equation}

\begin{equation}\label{14}
\varepsilon_{5} B_{\varphi}(\eta)-n\eta\frac{d B_{\varphi}(\eta)}{d\eta}+\frac{d}{d\eta}(v_{r}(\eta)B_{\varphi}(\eta))=0,
\end{equation}

\begin{equation}\label{15}
\varepsilon_{4} B_{z}(\eta)-n\eta\frac{d B_{z}(\eta)}{d\eta}+\frac{1}{\eta}\frac{d}{d\eta}(\eta v_{r}(\eta)B_{z}(\eta))=0,
\end{equation}

\begin{eqnarray}\label{16}
\frac{1}{\gamma-1}(\varepsilon_{6}P(\eta)-n\eta\frac{P(\eta)}{d\eta}+v_{r}(\eta)\frac{dP(\eta)}{d\eta})\nonumber\\
+\frac{\gamma}{\gamma-1}\frac{P(\eta)}{\eta}\frac{d}{d\eta}(\eta v_{r}(\eta))+R(\eta)^{2-\nu}P(\eta)^{\nu}=0,
\end{eqnarray}
where $\varepsilon_{1}$ - $\varepsilon_{7}$, and $n$ are obtained as follows,
\begin{eqnarray}\label{17}
\varepsilon_{1}&=&-2, \hspace{2mm} 2\varepsilon_{2}=2\varepsilon_{7}=\varepsilon_{3}=\frac{1}{\nu-1},\nonumber\\
n&=&\frac{2\nu-1}{2(\nu-1)}, \hspace{2mm} 2\varepsilon_{6}=\varepsilon_{4}=\varepsilon_{5}=\frac{3-2\nu}{\nu-1}.
\end{eqnarray}
Now the equations (\ref{10})-(\ref{16}) can be solved by using the fourth-order Runge-kuttta method which is considered in the next section.

\section{Initial boundary conditions and numerical solution}
\begin{figure}[th]
\begin{center}
\includegraphics[scale=.25]{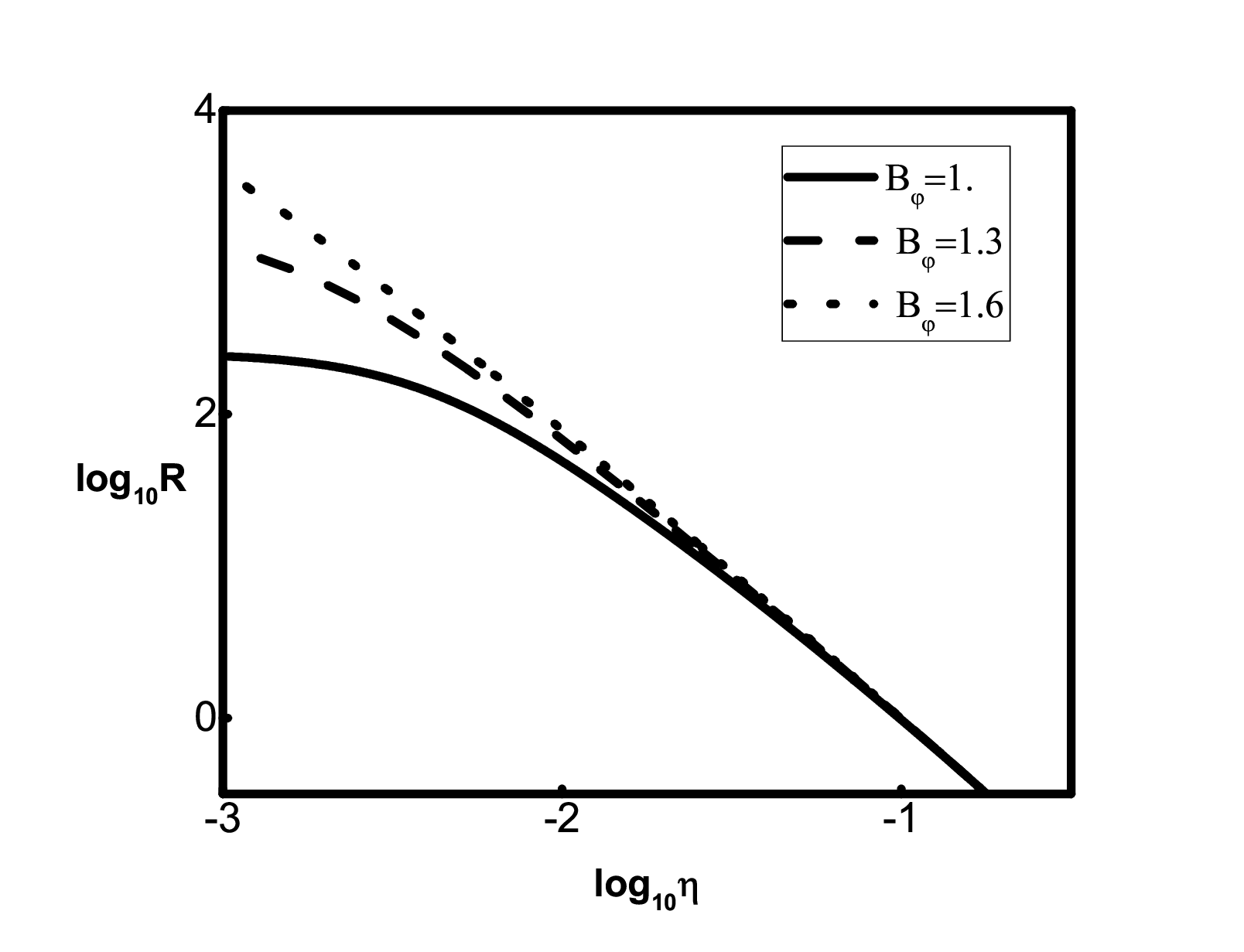}
\caption{Density profile corresponding to $\gamma=1.66$, and different initial value for azimuthal component of magnetic field.}
\label{fig1}
\end{center}
\end{figure}

In this section, we will use the observed outer conditions for filamentary clouds to integrate the equations from outside to inside of the cloud. We will study behavior of physical quantities. According to the observation of filamentary clouds, density is higher in the inner regions than outer regions. The outer density of filamentary cloud's is like ISM density. For the reason we will choose  $n=10 \hspace{2mm}cm^{-3}$ as typical density's value. We expect that this value increases in the center (Hanawa 1996). Density in the center is about  $n=10^{4} \hspace{2mm}cm^{-3}$ as typical density's value (Li and Goldsmith 2012; Henshow et al 2013). According to the boundary conditions and free parameter $\nu$, we can solve a set of ordinary equations. We can consider $\nu=2.4$ for a molecular cloud and determine other parameters with respect to that (Goldsmith 2001). In the Fig. \ref{fig1} we can see that density increases from outer to inner regions and is of order $n\approx10^{4} \hspace{2mm}cm^{-3}$ in agreement with recent work of Ysard et al. (2013). Shadmehri (2005) and Chapman et al. (2011) show that density increased with increasing of toroidal magnetic field component, in agreement with our result which illustrated by the Fig. \ref{fig1}.

According to the observation, the magnetic field has various shapes in the molecular cloud and makes different angels with filamentary cloud's axis in outer region (Planck Collaboration 2016). Tilley and Pudritz (2003) supposed that magnetic field in outer region in only azimuthal. Shadmehri (2005) supposed that the azimuthal component is dominate in the outer region, we suppose (in agreement with Wareing et al. 2016) that primary axial component of the magnetic field is smaller than  azimuthal magnetic field for outer region. Also, magnetic field in molecular cloud is of order $\mu G$ (Khesali et al 2014). Collapse velocity in outer region is less than 3 km/s. So we choose $v_{\varphi}=0.001$  and $v_{r}=-0.5$  as the dimensionless collapse velocity in outer region. Also we suppose that $\gamma=1.66$.

\begin{figure}[th]
\begin{center}
\includegraphics[scale=.25]{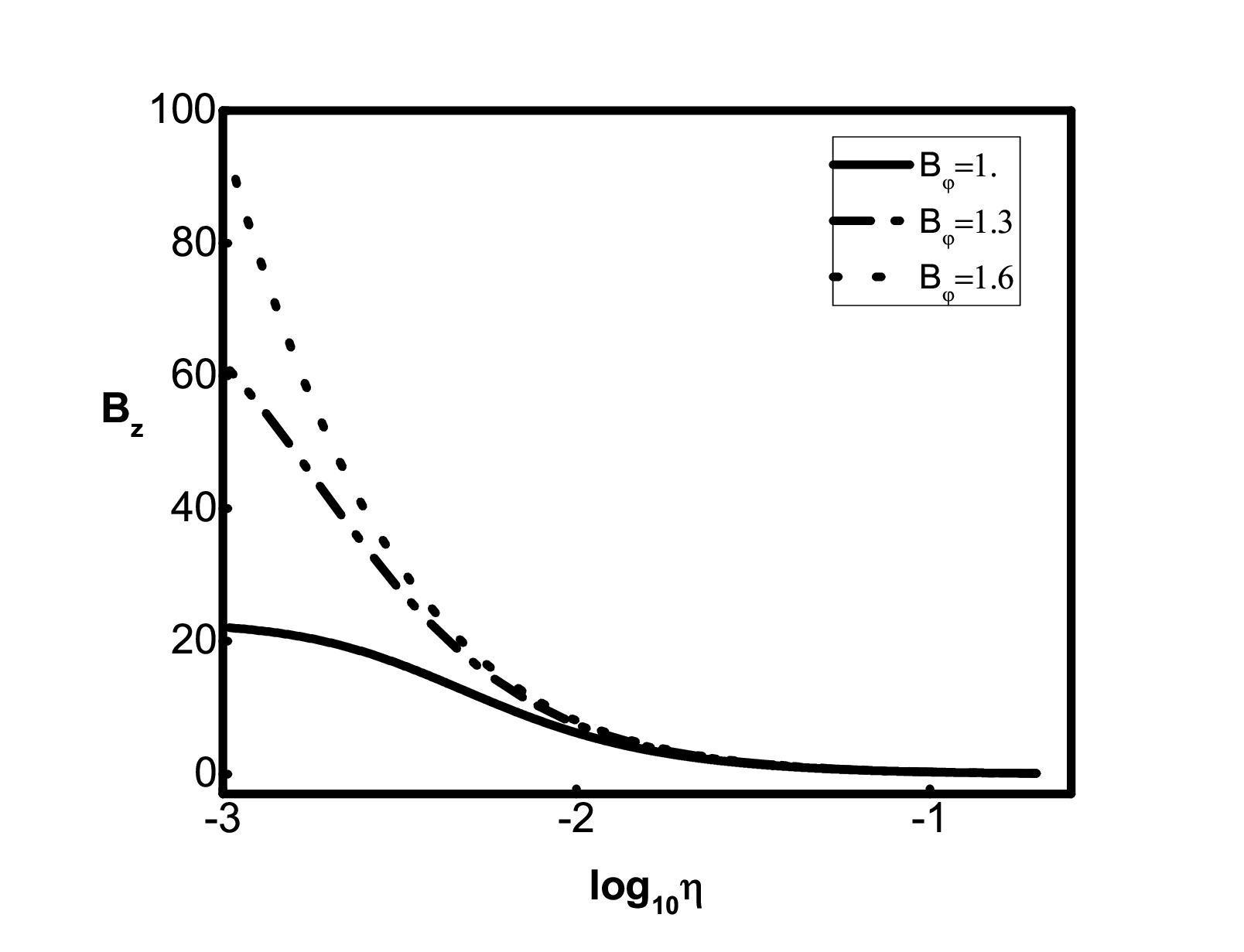}
\caption{Axial magnetic field corresponding to $\gamma=1.66$, and different initial value for azimuthal component of magnetic field.}
\label{fig2}
\end{center}
\end{figure}

\begin{figure}[th]
\begin{center}
\includegraphics[scale=.25]{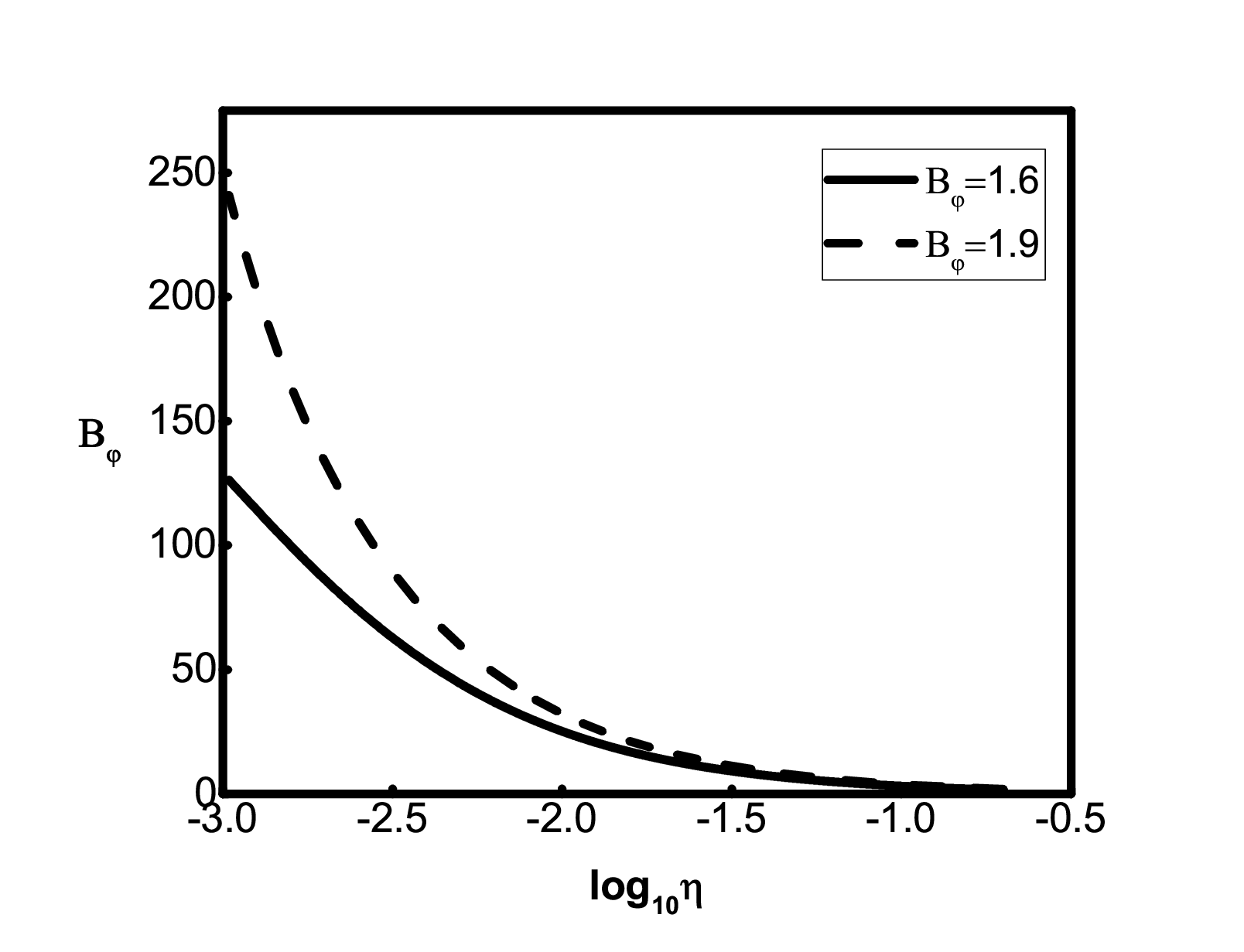}
\caption{Azimuthal magnetic field corresponding to $\gamma=1.66$, and different initial value for azimuthal component of magnetic field.}
\label{fig3}
\end{center}
\end{figure}

We show variation of axial and azimuthal magnetic fields by Figs. \ref{fig2} and \ref{fig3}, It is clear that value of magnetic field increased at center in agreement with previous works (Tilley and Pudritz 2003, Shadmehri 2005, Fiege and Pudritz 2000). Also we show that increasing initial azimuthal magnetic field in outer edge of cloud causes to increasing central magnetic field components. It is completely coincide with the Fig. \ref{fig1} which show magnetic field freezing of filament.

\begin{figure}[th]
\begin{center}
\includegraphics[scale=.25]{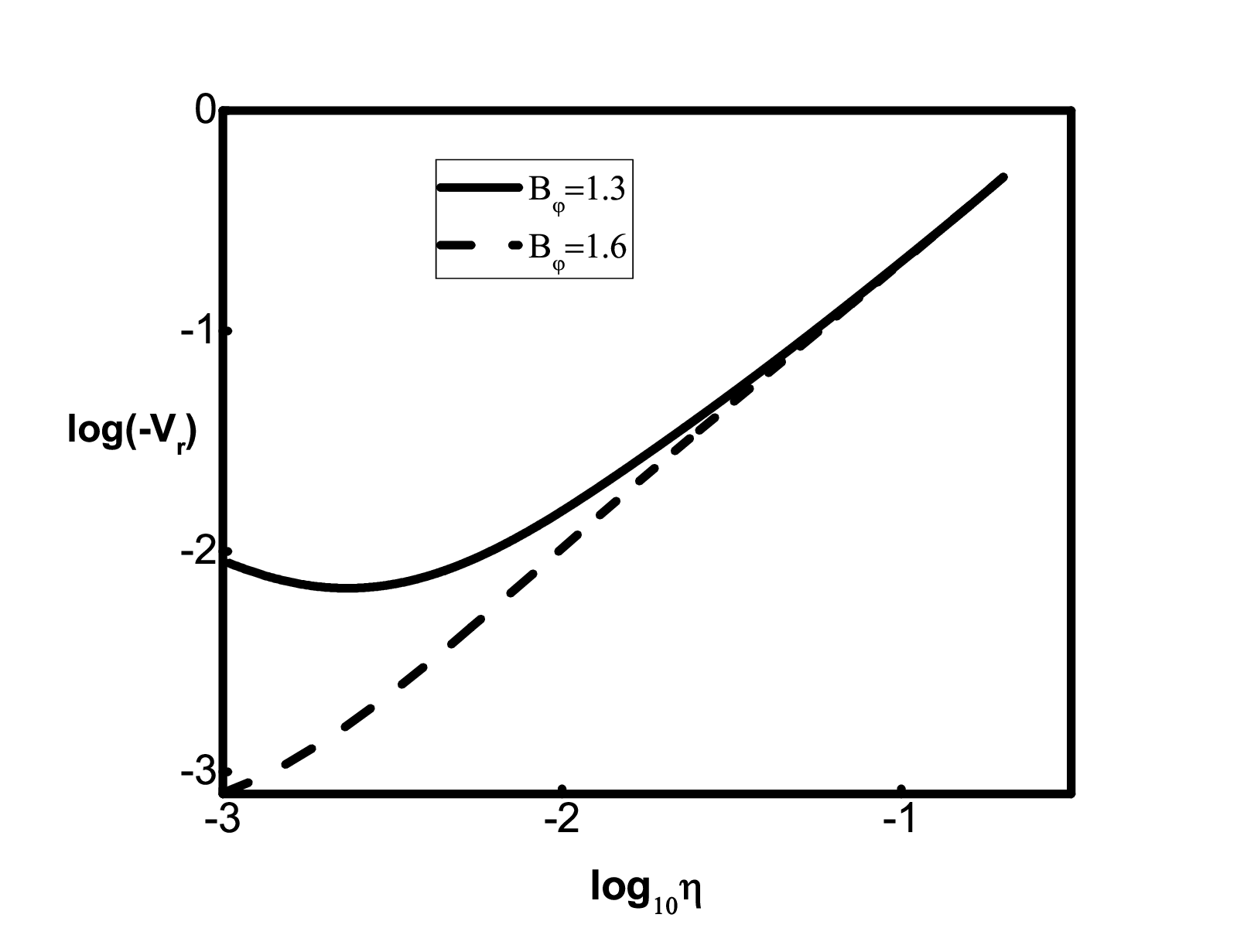}
\caption{radial velocity corresponding to $\gamma=1.66$, and different initial value for azimuthal component of magnetic field.}
\label{fig4}
\end{center}
\end{figure}

Previous works has been shown that magnetic field help to avoid gravitational collapse (Shadmehri 2005, K\"{o}rtgen and Banerjee 2015, Nakajima and Hanawa 1996). In agreement with mentioned work we show in the Fig. \ref{fig4} that radial velocity decreased with magnetic field in the center which is indeed magnetic braking effect.

\begin{figure}[th]
\begin{center}
\includegraphics[scale=.25]{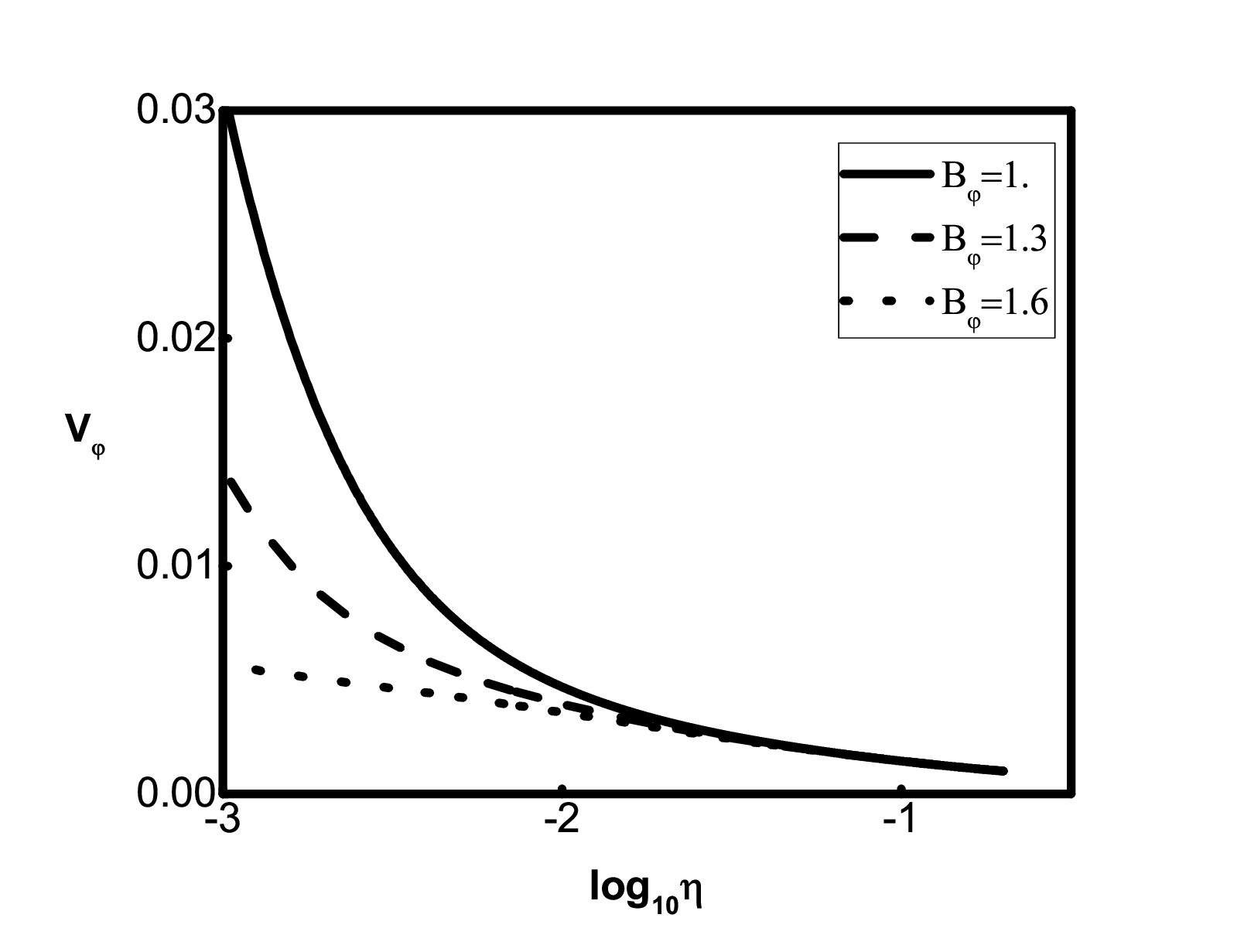}
\caption{Azimuthal velocity corresponding to $\gamma=1.66$, and different initial value for azimuthal component of magnetic field.}
\label{fig5}
\end{center}
\end{figure}

In the Fig. \ref{fig5} we can see that azimuthal velocity increased at center, but increasing of magnetic field reduced radial velocity due to the magnetic braking.

\begin{figure}[th]
\begin{center}
\includegraphics[scale=.25]{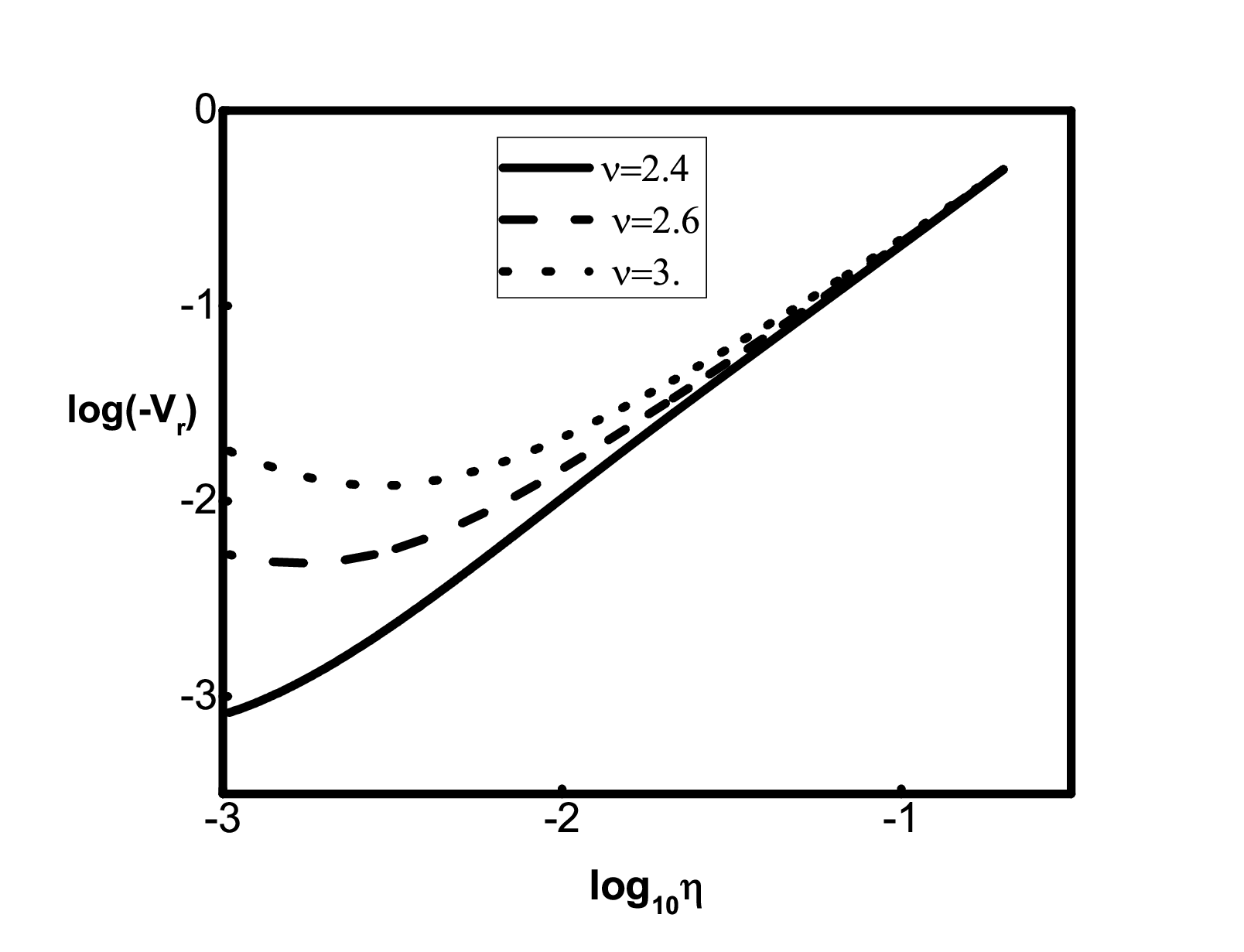}
\caption{radial velocity corresponding to $\gamma=1.66$, and different initial value for $\nu$.}
\label{fig6}
\end{center}
\end{figure}

\begin{figure}[th]
\begin{center}
\includegraphics[scale=.25]{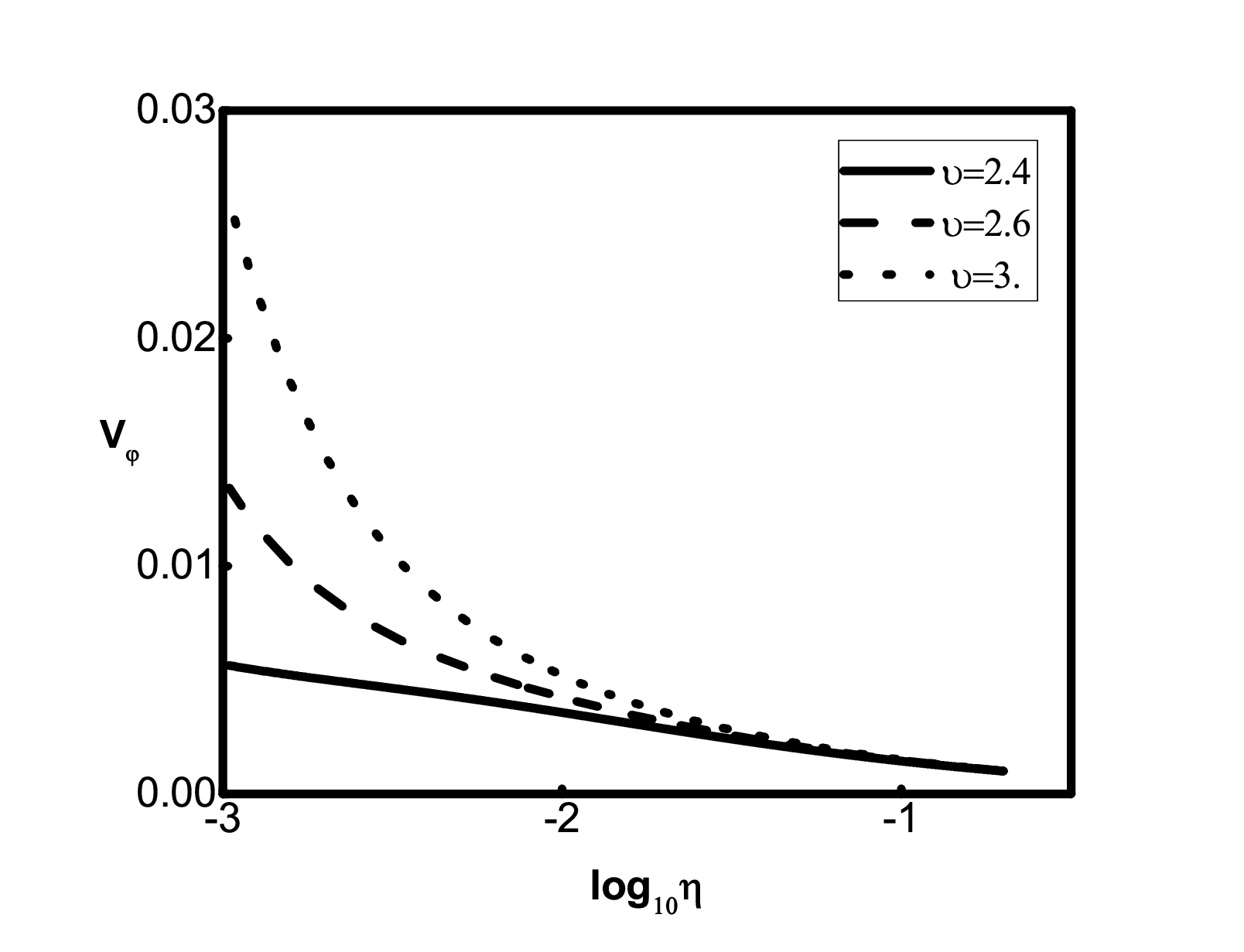}
\caption{Azimuthal velocity corresponding to $\gamma=1.66$, and different initial value for $\nu$.}
\label{fig7}
\end{center}
\end{figure}

We can see that when more energy is released from the filament the in fall velocity is increased as expected in the filament collapsing process. In that case in the Figs. \ref{fig6} and \ref{fig7} we see that increasing $\nu$ increases the value of radial and azimuthal velocities.

The Figs. 8 and 9 show effect of increasing energy release from the system on the density and magnetic field in agreement with the previous work (Khesali et al 2014).

\begin{figure}[th]
\begin{center}
\includegraphics[scale=.25]{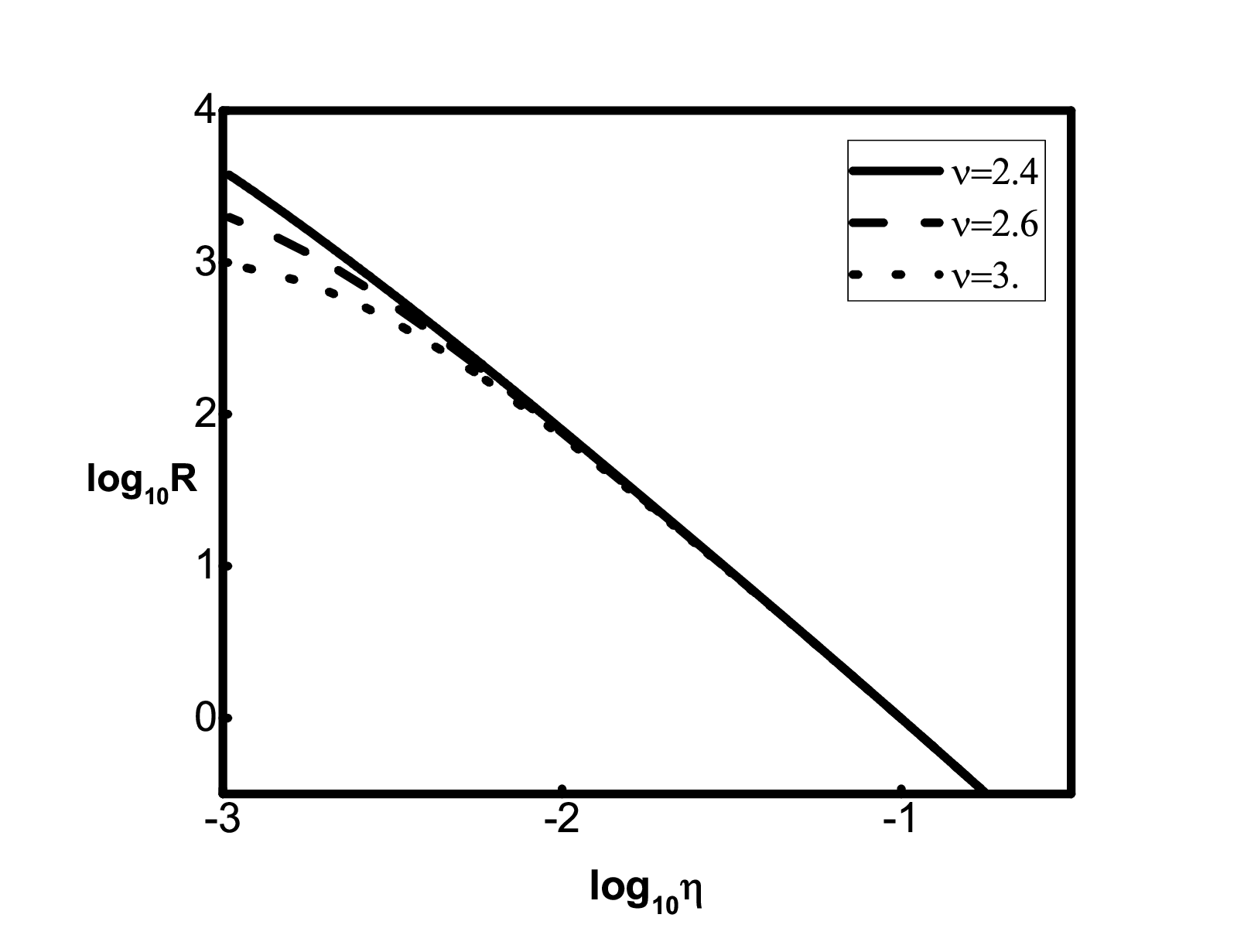}
\caption{Density profile corresponding to $\gamma=1.66$, and different initial value for $\nu$.}
\label{fig8}
\end{center}
\end{figure}

\begin{figure}[th]
\begin{center}
\includegraphics[scale=.25]{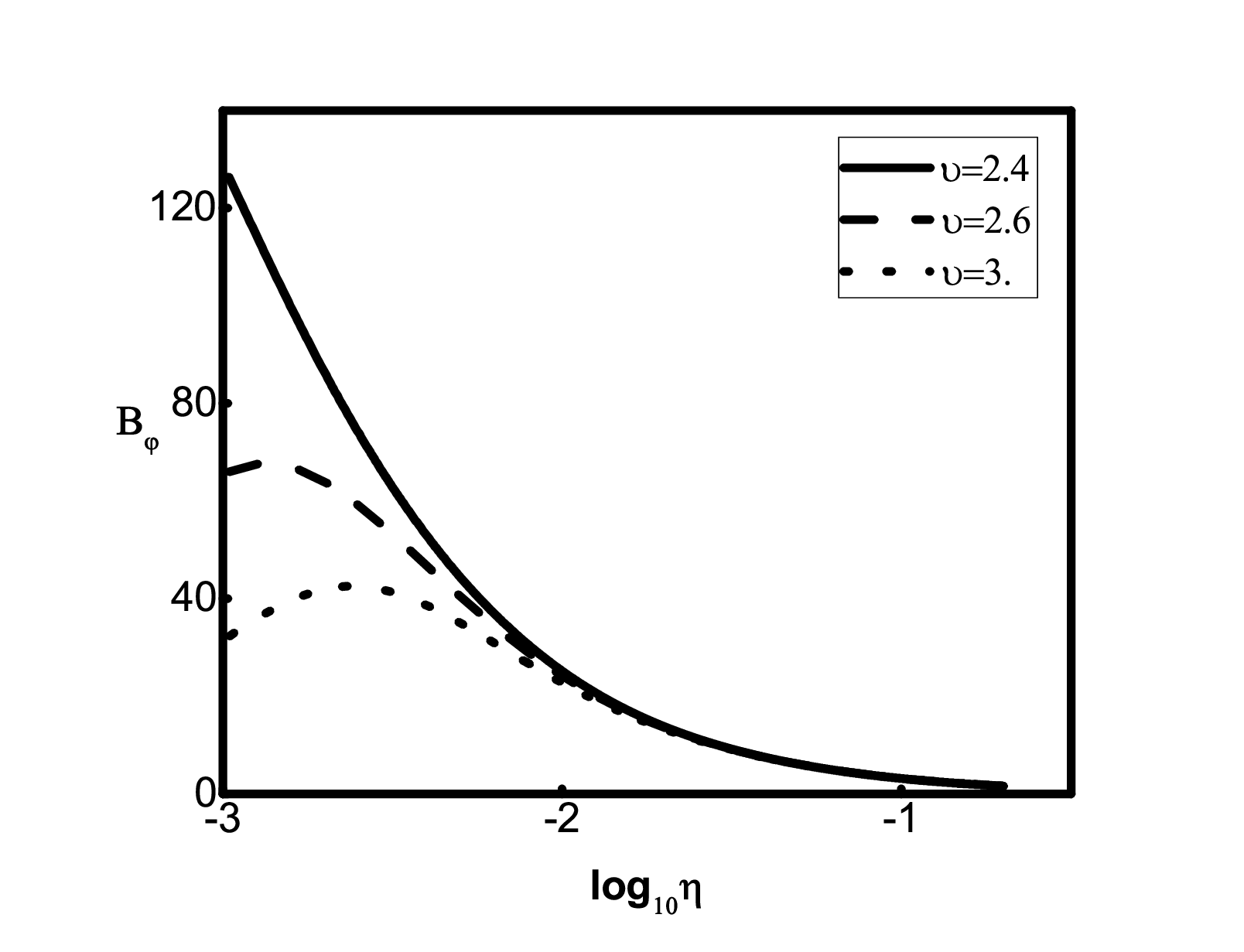}
\caption{Azimuthal magnetic field corresponding to $\gamma=1.66$, and different initial value for $\nu$.}
\label{fig9}
\end{center}
\end{figure}

In the Figs. 10 we draw density profile in terms of $z$-direction at the constant radius, and show that density profile affected by variation of magnetic field. It is illustrated that increasing of magnetic field decreases change of density, so it help to keep filament. Hence we can see less fragmentation in presence of magnetic field at the early stage.

\begin{figure}[th]
\begin{center}
\includegraphics[scale=.25]{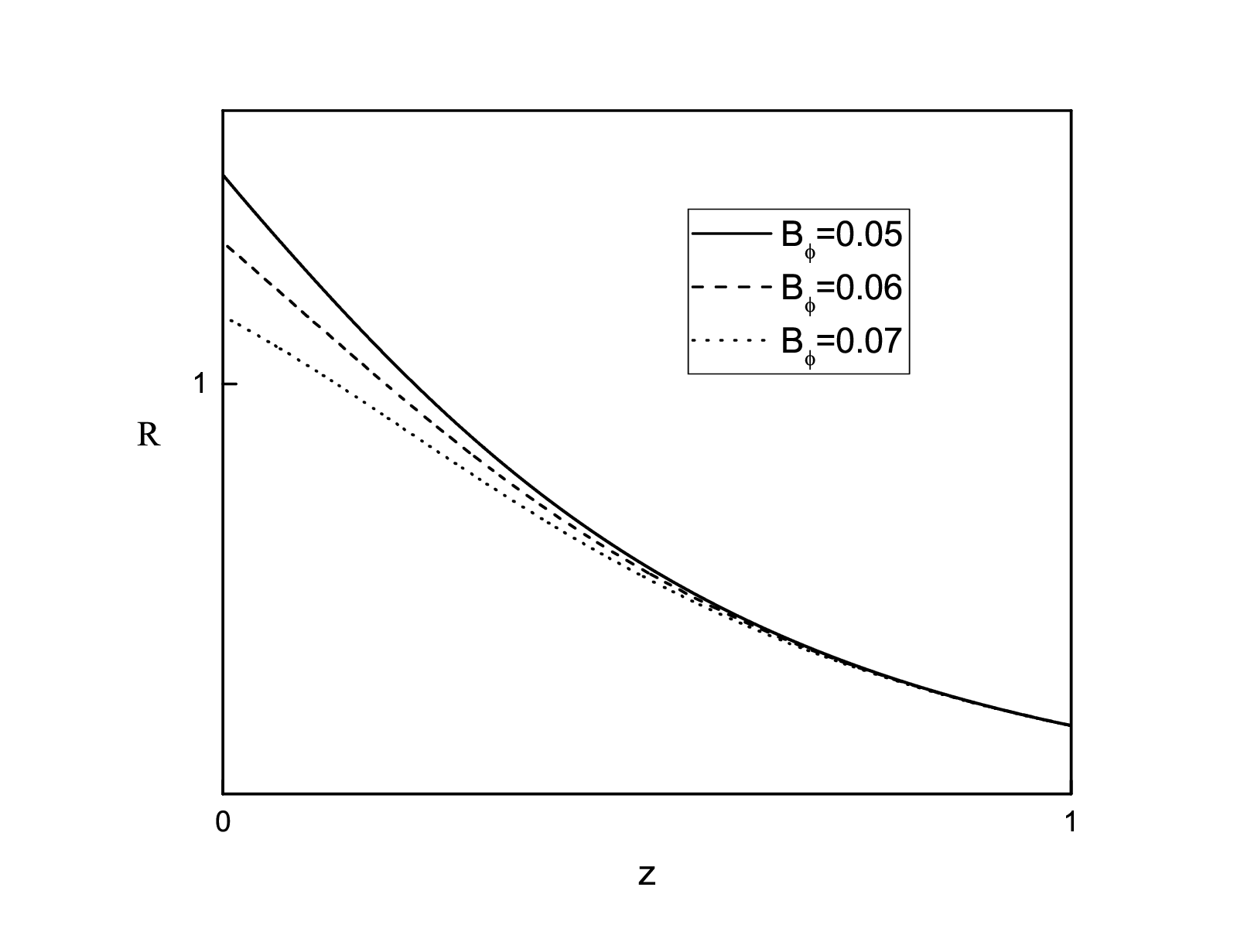}
\caption{Density profile corresponding to $\gamma=1.66$ with constant radius, and different initial value for azimuthal component of magnetic
field.}
\label{fig10}
\end{center}
\end{figure}

\section{Conclusion}
In this paper, we studied rotational filamentary molecular cloud's collapse in presence of magnetic field at the early stage.  We studied filamentary cloud's dynamic of the early stage with azimuthal and axial fields without introducing a function for magnetic field and density functions. Previous studies and observational data such as Arzoumanian et al. (2011) indicated that the filament density decreased with radius and our study also confirms these results as illustrated by the Fig. 1. Also, Kirk et al. (2015) has been studied the effect of magnetic fields in filamentary structure using the simulation method. We know that magnetic field has important role in the formation of the filament. Also, density increases with increasing of initial azimuthal magnetic field (Fig. 1).  Values of azimuthal and axial magnetic fields is larger in center as illustrated by Fig. 2 and Fig. 3, which show freezing of magnetic field. There are several models to formation of filament like Padoan et al. (2001) and Nakamura (2008) where importance of magnetic field have been studied. For example, K\"{o}rtgen and Banerjee (2015) have been shown that star formation due primary magnetic field heavily delayed or suppressed. Therefore, magnetic field can reduces radial collapse velocity which verified with the Figs. 4 of our work in agreement with other works like Shadmehri (2005) and  Khesali et al. (2014). Difference of our new results with Khesali et al. (2014) is presence of filament fragment due t rotational motion.
Behavior of radial and azimuthal velocity plotted in Fig. 4 and Fig. 5. In the Fig. 4 we can see that radial velocity is smaller at center to yields zero in agreement with observation (K\"{o}rtgen and Banerjee 2015). In the Fig. 5 we can see that azimuthal velocity is larger at center interpreted as magnetic braking. It means that bigger value of magnetic field help to stability of cloud. We know that rotational velocity is small at the early stage. It is clear that rotational velocity affected by the magnetic field. We shown that presence of magnetic field reduces value of rotational parameter and yield to more stability and this confirmed by several observations. In Figs. 6, 7, 8 and 9 effect of cooling function on the physical system parameters have been shown. In the Figs. 6 and 7 we can see that radial and azimuthal velocities increased by exit of energy from the system.  The Figs. 8 and 9 increasing exit energy effect on the density and magnetic field have been shown respectively.\\\\

\end{document}